%% file: chaptercaldarelli.tex
\documentclass[graybox]{svmult}

\usepackage{mathptmx}       
\usepackage{helvet}         
\usepackage{courier}        
\usepackage{type1cm}        
\usepackage{url}         
\usepackage{graphicx}        
\usepackage{multicol}        
\usepackage[bottom]{footmisc}
\usepackage{amssymb}    

\makeindex             

\begin{document}

\title*{Self-Organization and Complex Networks}
\author{Guido Caldarelli and Diego Garlaschelli}
\institute{Guido Caldarelli \at Centre SMC CNR-INFM, Dipartimento di Fisica 
Universit\`a ``Sapienza'', Piazzale A. Moro 5 00185 Roma, Italy, \email{Guido.Caldarelli@cnr.it}
\and Diego Garlaschelli \at Dipartimento di Fisica, Universit\`a di Siena, Via Roma 56, 53100 Siena, Italy. \email{Garlaschelli@unisi.it}}
%
%
\maketitle


\abstract{
In this chapter we discuss how the results developed within the theory of fractals and Self--Organized Criticality (SOC) can be fruitfully exploited as ingredients of adaptive network models. 
In order to maintain the presentation self--contained, we first review the basic ideas behind fractal theory and SOC. We then briefly review some results in the field of complex networks, and some of the models that have been proposed. Finally, we present a self--organized model recently proposed by Garlaschelli et al. [\emph{Nat. Phys.} {\bf 3}, 813 (2007)] that couples the fitness network model defined by Caldarelli et al. [\emph{Phys. Rev. Lett.} {\bf 89}, 258702 (2002)] with the evolution model proposed by Bak and Sneppen [\emph{Phys. Rev. Lett.} {\bf 71}, 4083 (1993)] as a prototype of SOC. Remarkably, we show that the results obtained for the two models separately change dramatically when they are coupled together. This indicates that self--organized networks may represent an entirely novel class of complex systems, whose properties cannot be straightforwardly understood in terms of what we have learnt so far.
}

\section{Introduction}
\label{sec:1}
Several important results on both the empirical characterization and the theoretical modelling of complex networks have been achieved in the last decade \cite{Calda,Calda2,DM03,siam,AB01}. Among the factors that have rendered this fast progress possible, one should surely acknowledge the unprecedented possibility to digitally store, and computationally analyse, huge datasets documenting the large--scale organization of biological, technological, and socio--economic systems. This has determined an empirically well--grounded problem of information extraction from a new form of data, where many units (vertices) are mutually interconnected by links (or edges), requiring novel paradigms for the identification of relevant patterns, and possibly regularities. 
A second reason is surely the scientific awareness, steadily grown during at least the last three decades, of the ubiquitous presence in nature of collective and emergent phenomena resulting from the interaction of many units within a complex system. In particular, the developments achieved within the broad fields of statistical physics, nonlinear dynamics, critical phenomena, fractal geometry, spin glasses, and many--body theory have contributed to the formation of a modern and interdisciplinary perspective, whose major focus is the (often unexpected) role of the interactions between constituents, rather than the individual details of the latter. 
Within this research field, whose boundaries are rather blurred, a diverse set of tools to handle the complexity of heterogeneous systems was developed. 
When the empirically--driven pressure towards the understanding of networks built up, the scientific community was faced with the possibility, and the challenge, to apply these tools to a genuinely new problem. 
As a result, some universal features across different real--world networks were identified, and theoretical models were proposed to reproduce and interpret them. At the same time, the scientific horizon extended even further, since a complete framework was not there to tackle the problem yet. Indeed, a satisfactory and unified approach to complex networks is still lacking, and this exciting field continues to attract the interest of a large community of scientists extending across different disciplines.\\

Broadly speaking, the main lines of research on networks that have been traced in the last decade are: \emph{i)} the definition and the empirical analysis of the static topological properties of networks; \emph{ii)} the modelling of (either static or growing) network formation; \emph{iii)} the effects that the topology has on various dynamical processes taking place on networks.
Some useful references  \cite{Calda,Calda2,DM03,siam,AB01} present reviews of these results. More recently, a few attempts to provide a unified approach to the problem have been proposed, exploiting the idea that these aspects of networks should in the end be related to each other. In particular, it has been argued that the complexity of real--world networks is in the most general case the result of the interplay between topology and dynamics. While most studies have focused either on the effects that topological properties have on dynamical processes, or on the reverse effects that vertex--specific dynamical variables have on network structure, it has been suggested that one should consider the mutual influence that these processes have on each other.
This amounts to relax the (often implicit) hypothesis that dynamical processes and network growth take place at well separated timescales, and that one is therefore allowed to consider the evolution of the fast variables while the slower ones are quenched. 
Remarkably, one finds that the feedback between topology and dynamics can drive the system to a steady state that differs from the one obtained when the two processes are considered separately \cite{natphys}. These results imply that adaptive networks generated by this interplay may represent an entirely novel class of complex systems, whose properties cannot be straightforwardly understood in terms of what we have learnt so far.\\

In what follows we shall review our contribution to this line of research. In particular, we shall present a self--organized model  \cite{natphys} where an otherwise static model of network formation driven by vertex \emph{fitness}  \cite{fitness} is explicitly coupled to an extremal dynamics process  \cite{BS} providing an evolution rule for the fitness itself. In order to highlight the novel phenomena that originate from the interplay between the two mechanisms, we first review the main properties of the latter when considered separately. 
In section \ref{sec:2} we recall some aspects of scale invariance and Self--Organized Criticality (SOC), and in particular the biologically--inspired Bak--Sneppen model  \cite{BS} where the extremal dynamics for the fitness was originally defined on static graphs. In section \ref{sec:3} we briefly review complex networks and in particular the so--called fitness model of network formation  \cite{fitness}, where the idea that network properties may depend on some fitness parameter associated to each vertex was proposed. Finally, in section \ref{sec:4} we present the self--organized model obtained by coupling these mechanisms.
The order of the presentation is also meant to highlight the fruitful synthesis that, as we have already mentioned, has originated by the application of ideas inherited by the previous understanding of complex systems to networks.

\section{Scale invariance and self--organization}
\label{sec:2}
Self--similarity, or fractality, is the property of an object whose subparts have the same shape of the whole. 
At first, self--similarity appeared as a peculiar property of a limited class of objects. Only later, due to the activity of Benoit Mandelbrot \cite{panza1,panza2}, it turned out that examples of fractal structures (even if approximate due to natural cutoffs) are actually ubiquitous in nature. 
Indeed, in an incredible number of situations the objects of interest can be represented by self--similar structures over a large, even if finite, range of scales. Examples include commodity price fluctuations \cite{panza1}, the shape of coastlines \cite{panza2}, the discharge of electric fields \cite{NPW84}, the branching of rivers \cite{RR96}, deposition processes \cite{BB84}, the growth of cities  \cite{BB94}, fractures \cite{MPP84}, and a variety of biological structures  \cite{scalinginbiology}.

\subsection{Geometric fractals}
Due to this ubiquity, scientists have tried to understand the possible origins of fractal behaviour. 
The first preliminary studies have focussed on mathematical functions built by recursion (Koch's snowflake, Sierpi\'nski triangle and carpet, etc.). Based on these examples, where self--similar geometric objects are constructed iteratively, mathematicians introduced quantities in order to distinguish rigorously between fractals and ordinary compact objects.\\

For instance, one of the simplest fractals defined by recursion is the Sierpinski triangle, named after the Polish mathematician Waclaw Sierpi\'nski who introduced it in 1915 \cite{Striangle}.
\begin{figure}
\sidecaption
\includegraphics[scale=.35]{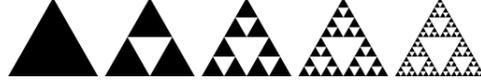}
\caption{First steps in the iteration procedure defining the Sierpinski triangle.}
\label{fig0}       
\end{figure}
When the procedure shown in Fig.\ref{fig0} is iterated an infinite number of times, one obtains an object whose empty regions extend at any scale (up to the maximum area delimited by the whole triangle). It is therefore difficult to measure its area 
in the usual way, i.e. by comparison with another area chosen as the unit of measure.
A way to solve this problem is to consider a limit process not only for the generation of the fractal, but also for the measurement of its area. 
Note that at the first iteration we only need three triangles of side length $1/2$ to cover the object (while for the whole triangle we would need four of them). At the second iteration we need nine covering triangles of side $1/4$ (while for the whole triangle we would need sixteen of them).
In general, for a compact triangle the number of triangles needed grows quadratically as we reduce the size of the covering triangles. The (scale--dependent) number of objects required to cover a fractal is at the basis of the definition of the {\bf fractal dimension} $D$.
Formally, if $N(\epsilon)$ is the number of $D_E$-dimensional volumes of linear size $\epsilon$ required to cover an object embedded in a metric space of Euclidean dimension $D_E$, then the fractal dimension is defined as  
\begin{equation} 
D=\lim_{\epsilon \rightarrow 0} \frac{\ln N(\epsilon)}{\ln 1/\epsilon},   
\label{eq:Fracd1} 
\end{equation} 
which approaches an asymptotic value giving a measure of the region occupied
by the fractal.\\

For a compact object the fractal dimension gives the same value as the 
Euclidean dimension $D_E$. Indeed, for the above compact triangle $D=D_E=2$.  
To see this, note that at the first iteration the number of necessary triangles 
is $4$ and $1/\epsilon$ is $2$, therefore $D= \frac{\ln 4}{\ln 2}=2$. At the next iteration $1/\epsilon$ is $4$ and the number of covering triangles is $16$ so that again $D= \frac{\ln 16} {\ln 4}=2$. Clearly, the same value of $D$ is found at all subsequent iterations, and therefore also in the limit $\epsilon \rightarrow 0$.
By contrast, for the Sierpi\'nski triangle it is easy to realise that at the $k$-th iteration the linear size of each covering triangle is $\epsilon=2^{-k}$ and that $N=3^k$ such triangles are needed. This implies
\begin{equation} 
D=\lim_{\epsilon \rightarrow 0} \frac{\ln N(\epsilon)}{\ln 1/\epsilon} =  
\frac{\ln 3}{\ln 2} \simeq 1.58496... 
\label{eq:Fracd2} 
\end{equation} 
Now we find that $D<D_E=2$. Therefore the fractal dimension measures the difference between the compactness of a fractal and that of a regular object embedded in a space of equal dimensionality. In the present example, $D$ is lower than $2$ because the Sierpinski triangle is less dense than a compact bidimensional triangle. $D$ is also larger than $1$ because it is denser than a one-dimensional object (a line). 
Note that the above formula can be rewritten in the familiar form of a power law by writing, for small $\epsilon$, 
\begin{equation} 
N(\epsilon) \propto {\epsilon}^{-D}
\end{equation} 
This highlights the correspondence between the geometry of a fractal and scale--invariant laws. 

\subsection{Self--Organized Criticality}
\label{subsec:2}
Despite their importance in characterizing the geometry of fractals, purely mathematical algorithms are not helpful in order to understand whether a few common mechanisms might be responsible for the fractal behaviour observed in so many different, and seemingly unrelated, real--world situations.
This has shifted the interest towards dynamical models.  
Indeed, open dissipative systems are in many cases associated with fractals for more than one reason. Firstly, attractors in the phase space of a nonlinear dynamical system can have a fractal geometry; secondly, their evolution can proceed by means of scale--invariant bursts of intermittent activity \cite{bursts} extending over both time and space. In general, these features are obtained when a driving parameter of the nonlinear dynamical system is set to a crossover value at which chaotic behaviour sets on. When this occurs, the nonlinear system is said to be at the ``edge of chaos''.
Another situation where self--similarity is observed is at the critical point of phase transitions. For instance, magnetic systems display a sharp transition from a high--temperature disordered phase, where microscopic spins point in random directions and generate no macroscopic magnetization, to a low--temperature ordered phase where almost all spins point in the same direction, determining a nonzero overall magnetization. Exactly at the critical transition temperature, spins are spatially arranged in aligned domains whose size is power--law distributed. This means that domains of all sizes are present, with a scale--invariant pattern.\\

In both cases, in order to explain the ubiquity of self--similar systems one should understand why they appear to behave as if their control parameter(s) were systematically fine--tuned to the critical value(s). 
This point led to the idea that feedback effects might exist, that drive the control parameter to the critical value as a spontaneous outcome of the dynamics. In this scenario, it is the system itself that evolves autonomously towards the critical state, with no need for an external fine--tuning. 
This paradigm is termed Self--Organized Criticality (SOC) (for a review 
see Ref. \cite{Jensen98} and references therein).
At a phenomenological level, SOC aims at explaining the tendency of open dissipative system to rearrange themselves in such a way to develop long--range temporal and spatial correlations. 
Why this happens is still a matter of debate, even if some authors claimed that this behaviour may be based on the minimization of some energy potential \cite{FO1,FO2,FO3}\footnote{Interestingly a similar claim 
has been made for networks as well \protect\cite{FS01}.}. Also, it has been proposed that a temperature--like parameter can actually be introduced for these systems  \cite{hot1,hot2}, and shown to lead to SOC only if fine--tuned to zero. This supports the hypothesis that SOC models are closely related
to ordinary critical systems, where parameters have to be tuned to their
critical value, the fundamental difference being the feasibility of this
tuning.\\ 

There are several examples of simplified models showing SOC, and most of them  have a common structure.
In practice, two classes of SOC models attracted many studies: the class of sandpile models \cite{BTW} and the class 
of models based on extremal dynamics such as the Bak--Sneppen \cite{BS} and Invasion Percolation \cite{WW83} models. In what follows we briefly review these examples.

\subsubsection{Sandpiles}\label{sec:sandpile}
One prototype is represented by {\em sandpile} models \cite{BTW}, a class of open dissipative systems defined over a finite box $\Lambda$ in a $d$--dimensional hypercubic lattice. In $d=2$ dimensions, one considers a simple square lattice. 
Any site $i$ of the lattice is assumed to store an integer amount $z_i$ of sand grains, corresponding to the height reached by the sandpile at that site.
At every time step one grain of sand is added on a randomly chosen site $i$, so that the height $z_i$ is increased by one. 
As long as $z_i$ remains below a fixed threshold, nothing happens
\footnote{Different functions of the height $z_i$ can be defined: for example the height
itself, the difference of height between nearest neighbours (first
discrete derivative of the height), the discrete Laplacian operator of
height (second discrete derivative), and so on.}.
But as soon as $z_i$ exceeds the threshold, the column of sand becomes unstable and ``topples'' on its nearest neighbours. Therefore the heights evolve according to
\begin{equation}
z_i \rightarrow z_i - \Delta_{ki}
 \end{equation}
where 
\begin{equation}
\label{eq:oper}
\Delta_{ki} =\left\{ 
\begin{array} {rl}
2d   &\quad k = i \\
-1   &\quad k\quad \mbox {\rm nearest neighbor of } i \\
0    &\quad \mbox {\rm otherwise.} 
\end{array}
\right. 
\end{equation}
This process is called {\em toppling}. As the neighbouring sites acquire new grains, they may topple in their turn, and this effect can propagate throughout the system until no updated site is active, in which case the procedures starts again with the addition of a new grain.
While the amount of sand remains constant when toppling occurs in the bulk, for topplings on the boundary sites ($i \in \partial \Lambda$) 
some amount of sand falls outside and disappears from the system. In the steady state of the process, this loss balances the continuous random addition of sand.\\ 

All the toppling events occurring between two consecutive sand additions are said to form an 
{\em avalanche}. One can define both a size and a characteristic time for an avalanche. 
The size of an avalanche can be defined, for instance, as the total number of
toppling sites (one site can topple more than once) or the total 
number of topplings (it is clear that these two definitions give more 
and more similar results as the space dimension increases). In order to define the lifetime of an avalanche, one must first define the unit timestep. The latter is the duration of the fundamental event defined by these two processes:
\begin{itemize}
\item{a set of sites becomes critical due to the previous toppling event;}
\item{all such critical sites undergo a toppling process, and the heights of their neighbours are updated.}
\end{itemize}
Then the lifetime of an avalanche can be defined as the number of unit timesteps 
between two sand additions. The fundamental result of the sandpile model is that at the steady state both the size $s$ and the lifetime $t$ of avalanches are characterized by power law distributions $P(s) \sim s^{-\chi}$, $Q(t) \sim t^{-\xi}$ \cite{BTW}. Therefore the model succeeds in reproducing the critical behaviour, often associated to phase transitions, with a self--organized mechanism requiring no external fine tuning of the control parameter. Note that the grain addition can be viewed as the action of an external field over the system. Similarly, the avalanche processes can be viewed as the response (relaxation) of the system to this field. The spatial correlations that develop spontaneously at all scales indicate that the system reacts macroscopically even to a microscopic external perturbation, a behaviour reminiscent of the diverging susceptibility characterizing critical phenomena. 

\subsubsection{The Bak--Sneppen model}\label{sec:bs}
A model that attempts to explain some key properties of biological evolution, even if with strong simplifications, is the Bak--Sneppen (BS) model  \cite{BS,flyvbjerg}. It is defined by the following steps:

\begin{itemize}
\item{$N$ species are arranged on the sites of a $1$-dimensional lattice (a chain, or a ring if periodic 
boundary conditions are enforced);}
\item{a \emph{fitness} value $x_i$ (sometimes interpreted as a fitness \emph{barrier}) is assigned to each species $i$, drawn randomly from a uniform distribution in the interval $[0,1]$;}
\item{the site with the lowest barrier and its nearest neighbours are updated: new random fitness values, drawn from the same uniform distribution on the unit interval, are assigned them.}
\end{itemize}
The basic idea behind the model is that the species with the lowest fitness is the one that is most likely to go extinct and replaced by a new one. Alternatively, the update is interpreted as a mutation of the least fit species towards an evolved species representing its descendant or offspring. Finally, one can interpret $x_i$ as the barrier against mutation for the genotype of species $i$: the higher the barrier, the longer the time between two 
modifications of the genetic code. The species with lowest barrier is therefore the first to evolve. 
In any case, the reason for updating the nearest neighbours 
is the same: the mutation of one species changes the state of all the interacting species (for instance, both predator and prey along the food chain). The effect of this change on the fitness of the nearest neighbours is not known \emph{a priori} (it may be beneficial or not), and is modelled as a random update of their fitness as well.\\

\begin{figure}
\sidecaption
\includegraphics[scale=.35]{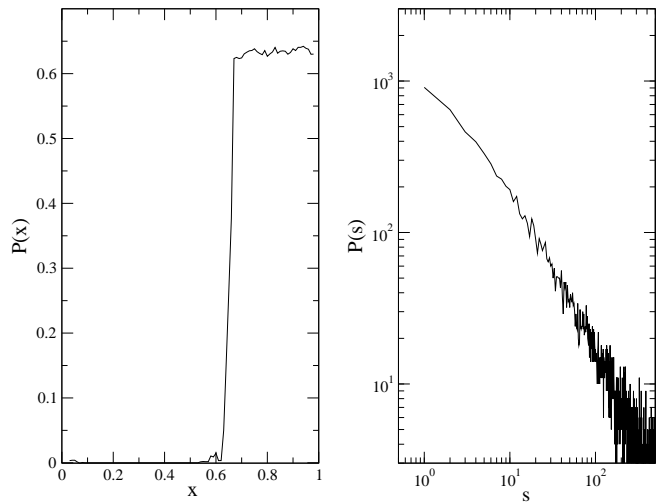}
\caption{Left: plot of the probability distribution of fitness values at the steady state in the Bak--Sneppen model with $500$ species. Right: the probability distribution $P(s)$ for the size of a critical $\tau$-avalanche.}
\label{fig1}       
\end{figure}

If the procedure described above is iterated, the system self--organizes to a critical stationary state 
in which almost all the barriers are uniformly distributed over a certain 
threshold value $\tau=0.66702 \pm 0.00008$   \cite{Grassberger} (see Fig.\ref{fig1}, left panel). In other words, the fitness distribution evolves from a uniform one in the interval $[0,1]$ to a uniform one in the interval $[\tau,1]$. 
In this model an (evolutionary) $x$-avalanche is defined as a causally connected sequence of mutations of barriers, all below a fixed value $x$. In this way the size of an $x$-avalanche 
is uniquely defined as the number of mutations between two consecutive configurations where all barriers are above $x$. 
For $x\approx \tau$ the avalanche distribution is a power law $P(s) \propto s^{-\chi}$ with an exponent $\chi=1.073 \pm 0.003$   \cite{Grassberger} (see Fig.\ref{fig1}, right panel).\\ 

The Bak--Sneppen model is a prototype mechanism generating fractal phenomena
as an effect of extremal dynamics  \cite{Dickman2000}. It also provides a possible explanation for the phenomena of mass extinctions observed in the fossil records  \cite{fossile}, some analyses of which have indicated that extinction sizes are power--law distributed. Rather than considering large--scale extinctions as triggered by external catastrophic events (such as meteorites or major environmental changes) and small--scale extinctions as caused by evolutionary factors, the model shows that a power--law distribution of extinction events may be interpreted as the outcome of a single internal macroevolutionary process acting at all scales.\\

The Bak--Sneppen model has been studied within a variety of different frameworks ranging from numerical simulation  \cite{Grassberger,DELOS1998},
theoretical analysis  \cite{Dorogovtsev2000},
renormalization group techniques  \cite{Marsili1994,Mikeska1997},
field theory  \cite{Paczuski1994}, mean-field approximations  \cite{flyvbjerg,Dickman2000} and probabilistic approaches (run time statistics)  \cite{Caldarelli2002,gabrie}. 
It has also been defined on higher--dimensional lattices and more general graphs, including complex networks  \cite{BS,flyvbjerg, gabrie,BSd,BSsw,BSsf,BSkim,BSkahng}, which are the subject of the next section. 
For a recent review on this model see ref. \cite{GD04} and references therein.
Being so well studied, the Bak--Sneppen model is ideal for studying the effects introduced by a feedback mechanism between fitness dynamics and topological restructuring. For this reason, it is at the basis of the adaptive model \cite{natphys} that we shall present in detail in section \ref{sec:4}.

\section{Complex networks}\label{sec:3}
Networks are encountered anywhere in nature  \cite{Calda,Calda2,DM03,siam,AB01}. For example, in biology they describe protein interactions, metabolic reactions, and gene expressions  \cite{droso,Giotetal2003,metabolic}. In the different context 
of ecology, food webs  \cite{webworld,nature} report predator--prey or host--parasite interactions, and taxonomic trees are used to classify different species  \cite{burlando1,burlando2,cecile}. Socio--economic systems display a strongly networked structure as well, for instance when considering the relationships between firms  \cite{shares} or trading countries  \cite{mywtw}. Technology produces network structures as well, the most striking evidences of which being the Internet and the WWW  \cite{FFF99,AH00,CMP00}.
During the last decade, it has been found that the overwhelming majority of real--world networks is characterized by nontrivial features, leading to the term ``complex networks''. As for the notion of ``complex systems'', a rigorous and/or widely accepted definition of complexity does not exist. Nonetheless, what is generally meant is that many topological properties of real networks are not easily reproduced by simple graph models. Quite surprisingly, these properties are often shared by networks of very different nature, suggesting common organization mechanisms.

\subsection{Network properties}
One of the widespread features observed in real networks is a scale--free distribution $P(k)\propto k^{-\gamma}$ for the degree $k$, representing the number of links emanating from a vertex. More formally, for an undirected network with $N$ vertices, the degree of each vertex $i$ can be expressed as
\begin{equation}\label{eq:degree}
k_i\equiv\sum_j a_{ij}
\end{equation}
where $a_{ij}=1$ if a link between $i$ and $j$ is there, and $a_{ij}=0$ otherwise. The empirical finding that $k_i$ is power--law distributed indicates  that even if the majority of vertices has a small number of neighbours, some of them (the ``hubs'') are connected to many vertices.\\

Another nontrivial property is the (anti)correlation between degrees of neighbouring vertices: vertices with a large value of the degree tend either to ``attract'' or to ``repel'' vertices with similar degree, a property known as \emph{assortativity} or \emph{disassortativity} respectively  \cite{Calda,siam}. This can be quantified by measuring the average degree of the nearest neighbours of a vertex $i$, defined as
\begin{equation}\label{eq:annd}
k^{nn}_i\equiv\frac{\sum_j a_{ij}k_j}{k_i}=\frac{\sum_{jk} a_{ij}a_{jk}}{\sum_j a_{ij}}
\end{equation}
and plotting it versus $k_i$. Assortative mixing corresponds to an increasing trend, while disassortative mixing corresponds to a decreasing trend of the resulting curve. In absence of correlations, a flat behaviour would be observed.\\

Another observed tendency is the presence of many more triangles (fully connected triples of vertices) than expected by chance, a feature denoted \emph{clustering}  \cite{Calda,siam}. For each vertex $i$, the clustering coefficient $c_i$ is defined as the fraction of links existing among its neighbours:
\begin{equation}\label{eq:c}
c_i\equiv
\frac{\sum_{jk}a_{ij}a_{jk}a_{ki}}{k_i (k_i -1)/2}=
\frac{\sum_{jk}a_{ij}a_{jk}a_{ki}}{\sum_{jk}a_{ij}a_{ki}}
\end{equation}
When plotted against $k_i$, for most real networks $c_i$ displays a decreasing trend, indicating the presence of \emph{hierarchy}. Unstructured networks would instead display a flat behaviour.
An average of $c_i$ over all vertices measures the overall probability that two vertices, both joined to a third one, are also connected to each other. This average clustering is found to be much larger than expected by chance.\\

High clustering is often combined with a small value of the average distance between pairs of vertices, and the term \emph{small world effect} is used to describe this combination  \cite{AB01}. Another property of interest is the existence in large networks of (sometimes overlapping) communities, modules, and ``rich clubs''  \cite{Calda,Calda2}. Besides their structural importance, these topological properties have a deep effect on the dynamical processes that take place on networks. Examples of processes whose dependence on the underlying network structure has been studied in detail include the spreading of epidemics  \cite{epidemics}, percolation  \cite{siam}, critical phenomena  \cite{critical}, the exchange of wealth  \cite{wealth1,wealth2}, and the sandpile  \cite{sandpileSF} and Bak--Sneppen models themselves  \cite{BS,flyvbjerg,gabrie,BSd,BSsw,BSsf,BSkim,BSkahng}.

\subsection{Network models}\label{sec:netmodels}
All these interesting properties are detected by comparing the topology, or the dynamical performance, of a network with a null model providing a randomized version of it. Graph models are therefore important benchmarks for understanding complex networks. Moreover, they are also used to test candidate mechanisms believed to be responsible for the onset of a particular topological feature, thus providing an insight into realistic network formation processes. 
The vast majority of theoretical models can be grouped in two broad classes. On one hand, one has static models with a fixed number of links and specified connection probabilities between them. This generates an ensemble of networks whose expected topological properties can be obtained analytically. The prototype of all static models is the random graph, that we shall briefly review in section \ref{sec:rg}. On the other hand, one has evolving models with a variable number of vertices and links, that grow under specified stochastic rules. The earliest example of these models is the one proposed by Barab\'asi and Albert  \cite{BA99}, and we shall present it in section \ref{sec:ba}. Most models proposed in the last decade are (often nontrivial) modifications of these two simple ones. For instance, in section \ref{sec:fm} we briefly review the fitness model, where the idea that the connection probability depends on some vertex--specific fitness has been introduced.
As we have anticipated in the Introduction, besides these two well established frameworks a third, more recent approach focuses on networks shaped by the interplay between dynamical processes defined on them and the readjustment of topology. Our main focus is exactly an example of such adaptive models, which shall be presented in detail separately in section \ref{sec:4}. 

\subsubsection{The random graph model}\label{sec:rg}
For an undirected network with $N$ vertices, the maximum possible number of edges (excluding self--loops) one can 
draw is given by $L_{max}=N(N-1)/2$. If all these edges are present, the graph is said to be ``complete''. At the opposite limit, if no edge is present, the graph is said to be ``empty''.
In between these two extremes, one can form instances of more or less dense networks by drawing each of the possible edges independently with a probability $p$. This defines the random graph model  \cite{AB01}, whose only parameter (besides $N$) is $p$.
The case $p=0$ recovers the empty graph, while the case $p=1$ yields the complete one.
The expected number (average $\langle\cdots\rangle$ over the ensemble of possible realisations) of edges in a random graph with probability $p$ is given by 
\begin{equation}\label{eq:l}
\langle L\rangle= p \frac{N(N-1)} {2} 
 \end{equation}
and the expected degree, which is the same for all vertices, is
\begin{equation}\label{eq:pn}
\langle k\rangle=p(N-1)\approx pN.  
\end{equation}
For $N$ large the correlations between the various degrees can be neglected (degrees are not independent in a finite graph), and the degree distribution $P(k)$ can be approximated by the probability that a single vertex has degree $k$. To obtain a vertex with degree $k$, we must have  
$k$ times a successful event whose probability is $p$, and $(N-1-k)$ times  
an unsuccessful event whose probability is $(1-p)$. 
Since this can happen in 
\begin{equation}
\left(\!\!\!\begin{array}{c} N-1 \\k 
\end{array}\!\!\!\right)  
=\frac{(N-1)!}{(N-1-k)!k!} 
\end{equation}
combinations, we have  
\begin{equation} 
P(k)= \left(\!\!\! \begin{array}{c} N-1 \\k \end{array}\!\!\! \right) 
p^k(1-p)^{N-1-k}  
\end{equation}
The distribution is automatically normalized since 
\begin{equation} 
\sum_{k=0}^{N-1} P(k)= \left[p+(1-p)\right]^{N-1}=1.  
\end{equation} 
The above binomial distribution is well approximated by a 
Poisson distribution in the limit $N\rightarrow \infty$ and $p 
\rightarrow 0$ (with $Np$ kept constant): 
\begin{equation} 
P(k)\approx \frac{(Np)^ke^{-pN}}{k!} =\frac{\langle k \rangle^k e^{-\langle k \rangle}}{k!}.  
\end{equation}  
where we have used eq.(\ref{eq:pn}). Thus the degree distribution of the random graph decays exponentially, and is well concentrated about the average value $\langle k\rangle$. This is in stark contrast with the scale--free behaviour of most real networks, characterized by the power--law tail of $P(k)$.\\

The expected value of the average nearest neighbours degree defined in eq.(\ref{eq:annd}) is the same for all vertices as well, and equals the average degree:
\begin{equation}
\langle k^{nn}\rangle=\frac{p^2(N-1)^2}{p(N-1)}=p(N-1)
\end{equation}
This means that, as expected, in the random graph no (dis)assortative mixing is present, and the degrees of neighbouring vertices are uncorrelated.\\

Similarly, for the expected value of the clustering coefficient defined in eq.(\ref{eq:c}) one finds
\begin{equation}
\langle c\rangle=\frac{p^3(N-1)(N-2)}{p^2(N-1)(N-2)}=p
\end{equation}
so that no hierarchical structure is present. Moreover, if the value of $p$ is chosen in such a way that the expected number of links in eq.(\ref{eq:l}) matches the empirically observed one, then the resulting value of $\langle c\rangle$ is much smaller that the observed average clustering coefficient.\\

One can also derive an upper bound for the average distance, by considering the \emph{diameter} $D$ (defined as the maximum distance between pairs of vertices).
Exploring the graph as in a breadth first search algorithm, one finds 
that if the number of first neighbours of a vertex is $\langle k \rangle$, and if the network is connected, then the number of vertices visited after $d$ steps must be approximately $\langle k \rangle^d$. The total number $N$ of vertices is reached in at most $D$ steps, so that
\begin{equation}
N \gtrsim \langle k \rangle ^D \quad \Rightarrow\quad D \lesssim \frac{\ln N}{\ln \langle k\rangle}.
\end{equation}
Therefore the average distance scales at most logarithmically with $N$, a feature which is consistent with the small values observed.\\
 
In summary, for random graphs 
\begin{itemize}
 \item no scale--free degree distribution is present;
 \item degrees of neighbouring vertices are uncorrelated;
 \item the clustering is too weak and not hierarchical;
 \item no small world effect is present, even if the average distance is small.
\end{itemize}
 
\subsubsection{The Barab\'asi-Albert model}\label{sec:ba}
The Barab\'asi-Albert model  \cite{BA99} is the prototype of evolving network models, where it is assumed that the system grows at any time step. 
Both the number of vertices and the number of edges increase with  
time, since new vertices enter the network and are assumed to connect to the pre--existing ones with a probability proportional to the degree of the latter ({\it rich-get-richer} mechanisms). This implies that newcomers 
establish their connections preferentially with vertices that already have  
a large degree. 
It is then clear that the two novel ingredients in this model of network formation are {\em growth} and {\em preferential attachment}. The main success of the model is that these two simple rules produce naturally scale--free networks with degree distribution $P(k)\propto k^{-\gamma}$ (where $\gamma=3$).\\

In order to derive this result, we rephrase the model quantitatively.
The initial ($t=0$) state consists of $N_0$ vertices and no link.
At each timestep $t$ a new vertex attached to $m_0$ new edges enters the system. The loose ends of these $m_0$ edges connect to $m_0$ pre--existing vertices, chosen with a probability $\Pi(k_i,t)$ proportional to their degree at time $t$: 
\begin{equation} 
\Pi(k_i,t)= \frac{k_i(t)}{\sum_{j} k_j(t)} 
\end{equation} 
This directly implies that the numbers of vertices and edges at time $t$ are given by
\begin{eqnarray} 
N(t)&=&N_0+t \nonumber \\ 
m(t)&=&\frac{1}{2}\sum_{j} k_i(t)=m_0t. 
\end{eqnarray}
Using a continuous--time approximation, one can write the time evolution of the degree $k_i$ by noting that its rate of increase is
\begin{equation} 
\frac {\partial k_i}{\partial t}=
m_0\Pi(k_i,t)=
m_0\frac{k_i(t)}{\sum_j k_j(t)}=
\frac{m_0 k_i(t)}{2m_0t}=  
\frac{k_i(t)}{2t}
\end{equation} 
The above differential equation can be solved using the initial condition $k(t_i)=m_0$, where $t_i$ is the time when vertex $i$ entered the network. The solution is
\begin{equation} 
k_i(t)=m_0\left(\frac t {t_i}\right)^{1/2}
\end{equation} 
showing that the degree grows with the square root of time.  
This relation allows us to compute the exponent of the degree distribution.   
The probability $P(k_i<k)$ that a vertex has a degree smaller 
than $k$ is $P(k_i<k)= P\left(t_i>\frac{m_0^2t}{k^2}\right)$.  
Since vertices enter at a constant rate, the distribution of their injection times is uniform between the initial time $t_i=0$ and the current time $t_i=t$. In this interval, $P(t_i)=1/N(t)=1/(N_0+t)$. 
This implies 
\begin{equation} 
P\left(t_i>\frac{m_0^2t}{k^2}\right) = 1-P\left(t_i\le\frac{m_0^2t}{k^2}\right)=
1-\frac{m_0^2t}{k^2} \frac{1} {(N_0+t)} 
\end{equation} 
from which we have  
\begin{equation} 
P(k)=\frac {\partial P(k_i<k)}{\partial k} = 
\frac{2m_0^2t}{(N_0+t)} \frac 1 {k^3}\propto k^{-3}  
\end{equation} 
Therefore, we find that the degree distribution is a power law with a 
value of the exponent $\gamma =3$.\\ 

This derivation highlights the difficulty, as compared with static models, of deriving exact results for growing networks, which are therefore often explored by means of numerical simulations. Despite this difficulty, a series of results have been derived for the model. We only list some of them by reporting that networks generated by the Barab\'asi-Albert model 
\begin{itemize}
 \item have power--law distributed degrees (as shown above);
 \item have no correlations between degrees of neighbouring vertices  \cite{siam};
 \item show a clustering larger than the random graph case \cite{FFH03,BP05};
 \item display the small--world effect \cite{BR01}.	
\end{itemize}

\subsubsection{The fitness model}\label{sec:fm}
A completely different approach to obtain self--similar networks is to extend in a suitable way the random graph model defined in section \ref{sec:rg}. In the latter, all vertices are assumed to be statistically equivalent, so unsurprisingly no heterogeneity emerges. 
By contrast, one can define a static model where heterogeneity is explicitly introduced at the level of vertices. In particular, Caldarelli et al. \cite{fitness} have proposed a model where each vertex $i$ ($i=1,\dots,N$) is assigned a \emph{fitness} $x_i$ drawn from a specified distribution $\rho(x)$. Then, each pair of vertices $i$ and $j$ is sampled, and a link is drawn between them with a fitness--dependent probability $p_{ij}=f(x_i,x_j)$.
The expected topological properties of the network can be easily computed in terms of $\rho(x)$ and $f(x,y)$ \cite{fitness,pastor,servedio}. For instance, the expected degree of vertex $i$ is
\begin{equation}\label{eq:fitk}
\langle k_i\rangle=\sum_j p_{ij}=\sum_j f(x_i,x_j) 
\end{equation}
For $N$ large, the discrete sum can be approximated by an integral. Thus the expected degree of a vertex with fitness $x$ is
\begin{equation}\label{eq:kx}
k(x)=N\int f(x,y)\rho(y)dy 
\end{equation}
where the integration extends over the support of $\rho(x)$.
If one consider the cumulative fitness distribution and the cumulative degree distribution defined as
\begin{equation}\label{eq:rho>}
\rho_>(x)\equiv\int_x^{+\infty}\rho(x')dx'\qquad P_>(k)\equiv\int_k^{+\infty}P(k')dk'
\end{equation}
then the latter can be easily obtained in terms of the former as 
\begin{equation}\label{eq:Prho}
P_>(k)=\rho_>[x(k)]
\end{equation}
where $x(k)$ is the inverse of the function $k(x)$ defined in eq.(\ref{eq:kx}).\\

Similarly, the expected value of the average nearest neighbours degree defined in eq.(\ref{eq:annd}) is 
\begin{equation}\label{eq:fitannd}
\langle k^{nn}_i\rangle=
\frac{\sum_j p_{ij}\langle k_j\rangle}{\langle k_i\rangle}=
\frac{\sum_{jk} p_{ij}p_{jk}}{\sum_j p_{ij}}
\end{equation}
and the expected value of the clustering coefficient defined in eq.(\ref{eq:c}) is 
\begin{equation}\label{eq:fitc}
\langle c_i\rangle=
\frac{\sum_{jk}p_{ij}p_{jk}p_{ki}}{\langle k_i\rangle(\langle k_i\rangle-1)/2}=
\frac{\sum_{jk}p_{ij}p_{jk}p_{ki}}{\sum_{jk}p_{ij}p_{ki}}
\end{equation}
As for eq.(\ref{eq:fitk}), the above expressions can be easily rephrased in terms of integrals involving only the functions $f(x,y)$ and $\rho(x)$, upon which all the results depend.\\

The constant choice $f(x,y)=p$ is the trivial case corresponding to a random graph, irrespectively of the form of $\rho(x)$. The simplest nontrivial choice can be obtained requiring that the fitness--dependent network has no degree correlations other that those introduced by the local properties alone. It can be shown that this requirement leads to the form  \cite{newman_origin,likelihood}
\begin{equation}
f(x,y)=\frac{zxy}{1+zxy}
\label{fermi}
\end{equation}
where $z$ is a positive parameter controlling the number of links. Apart for the so--called \emph{structural correlations} induced by the 
degree sequence   \cite{newman_origin,likelihood}, higher--order properties are 
completely random, as in the \emph{configuration model}   \cite{siam,maslov}. 
When $z<<1$, the above connection probability reduces to the bilinear choice
\begin{equation}
f(x,y)=zxy
\label{bilinear}
\end{equation}
In this case, a sparse graph is obtained where structural correlations disappear. Also, from eq.(\ref{eq:fitk}) one finds that $\langle k_i\rangle\propto x_i$. 
If one chooses a power--law fitness distribution $\rho(x)\propto x^{-\gamma}$, it is therefore clear that the degree distribution will have exactly the same shape: $P(k)\propto k^{-\gamma}$. In the more general case corresponding to eq.(\ref{fermi}), the same choice for $\rho(x)$ yields again a power--law degree distribution, with a cut--off at large degree values that correctly takes into account the requirement $k\le N$ for dense. Equation (\ref{fermi}) also generates disassortativity and hierarchically distributed clustering, both arising as structural correlations imposed by the local constraints. For sparse networks, corresponding to eq.(\ref{bilinear}), these correlations disappear.\\

Another interesting choice is given by
\begin{equation}
f(x,y)=\Theta(x+y-z)\qquad \rho(x)=e^{-x}
\label{eq:threshold}
\end{equation}
where $z$, which again controls the number of links, now plays the role of a positive threshold. This choice yields again a power--law degree distribution $P(k)\propto k^{-\gamma}$ (where now $\gamma=2$), anticorrelated degrees with $k^{nn}(k)\propto k^{-1}$, and hierarchically distributed clustering $c(k)\propto k^{-2}$ (times logarithmic corrections) \cite{fitness,pastor,servedio}. 
Remarkably, it has been shown that both eq.(\ref{fermi}) and eq.(\ref{eq:threshold}) are particular cases of a more general expression obtained by introducing a temperature--like parameter  \cite{temperature}. Equation (\ref{fermi}), with $\rho(x)\propto x^{-\gamma}$, corresponds to the finite--temperature regime, where the temperature can be reabsorbed in a redefinition of $x$ and $z$. By contrast, eq.(\ref{eq:threshold}) corresponds to the zero--temperature regime where the structural correlations disappear and the graph reaches a sort of ``optimized'' topology  \cite{temperature}. In all these cases, the average distance is small. In summary, for a series of reasonable choices the networks generated by the fitness model display
\begin{itemize}
 \item a scale--invariant degree distribution;
 \item correlations between neighbouring degrees;
 \item hierarchically distributed clustering;
 \item a small--world effect.	
\end{itemize}

\section{A self--organized network model}\label{sec:4}
As we have anticipated in the Introduction and in section \ref{sec:netmodels}, more recent approaches to the modelling of complex networks have considered the idea that the topology evolves under a feedback with some dynamical process taking place on the network itself (see for instance refs.  \cite{natphys,webworld,jain,ginestra,holyst,zanette,plos,kozma}). 
Among the various contributions, three groups have considered a possible connection with Self--Organized Criticality  \cite{natphys,ginestra,holyst}.\\

Bianconi and Marsili  \cite{ginestra} have defined a model where slow network growth, defined as the gradual addition of links between randomly chosen vertices, is combined to fast relaxation, defined as the random rewiring of links connected to congested (toppling) vertices. To avoid the collapse to a complete graph, dissipation is also introduced, allowing toppling nodes to lose all their links at a given rate. 
The outcomes of the model depend on the dissipation rate and on the probability density function for the toppling probabilities to be assigned at each vertex. 
A particular choice of these quantities drives the system to a stationary state characterized by a scale--free topology and a power--law distribution for toppling avalanches.\\

Fronczak, Fronczak and Holyst  \cite{holyst} have proposed a model where no parameter choice is required in order to drive the system to the critical region. They considered the sandpile dynamics defined in section \ref{sec:sandpile}, but where each vertex has a different critical height equal to its degree, as in other previous studies  \cite{sandpileSF}. In addition, they assumed that after an avalanche of size $A$, the $A$ ends of links in the network that have not been rewired for the longest time are rewired to the initiator of the avalanche. In this way, the avalanche area distribution and the degree distribution evolve in time, and at the stationary state become very similar and scale--free.\\

Garlaschelli, Capocci and Caldarelli  \cite{natphys} have introduced another fully self--organized model where the Bak--Sneppen dynamics defined in section \ref{sec:bs} takes place on a network whose topology is in turn continuously shaped by the fitness model presented in section \ref{sec:fm}. Remarkably, they find that the mutual interplay between topology and dynamics drives the system to a state characterized by scale--free distributions for both the degrees and the fitness values. These unexpected properties differ from what is obtained when the two models are considered separately. The rest of the chapter is devoted to a detailed description of this model.

\subsection{Motivation}\label{subsec:bscn}
We have already mentioned that the topology of a network affects dramatically the outcomes of dynamical processes taking place on it  \cite{Calda,Calda2,siam,AB01}. 
On the other hand, the idea behind the fitness model presented in section \ref{sec:fm} captures the empirically observed result  \cite{shares,mywtw,duygu} that the topology of many real networks is strongly dependent on some vertex--specific quantity.
Clearly, these results imply that in general one should consider the mutual effects that dynamics and topology have on each other. 
Unfortunately, the overwhelming majority of studies have instead considered the two processes separately, by postulating either a scenario where the topology evolves over a much longer timescale than the dynamics, or the opposite situation where the dynamical variables evolve much more slowly than the topology (and are therefore assumed fixed as in the fitness model itself). 
In cases when there is indeed such a sharp separation of timescales, these approaches are helpful. But in many cases the topological evolution and the dynamics may occur at comparable rates, in which case the decoupled approach gives no insight into the real process.
Moreover, even when the timescales are indeed well separated, it is clear that the variables involved in the slower of the two processes must be specified as external parameters, and \emph{ad hoc} assumptions must therefore be made. For instance, when considering the spreading of epidemics on a network one should assume an arbitrary fixed topology. Similarly, when a network is formed according to the fitness model, one should assume an arbitrary distribution for the fitness variables.\\

These motivations lead Garlaschelli et al. \cite{natphys} to define a self--organized model where \emph{ad hoc} specifications of any fixed structure, either in the topology or in the dynamical variables, are unnecessary. Rather, it is the interplay between dynamics and topology that autonomously drives the system to a stationary state. 
The choice of both the dynamical rule and the graph formation process was driven by the interest to highlight the novel effects arising uniquely by the feedback introduced between them. Therefore, two extremely well understood models where chosen. On one hand, the extremal fitness dynamics of the Bak--Sneppen model (see section \ref{sec:bs}), and on the other hand the fitness network model (see section \ref{sec:fm}).
As we have shown in section \ref{sec:fm}, the topology generated by the fitness model can be completely calculated for any distribution of the fitness values.
Similarly, the outcomes of the Bak--Sneppen model on several static networks are well studied  \cite{BS,flyvbjerg,gabrie,BSd,BSsw,BSsf,BSkim,BSkahng}. On a generic graph, each of the $N$ vertices is assigned a fitness value $x_i$, initially drawn from a uniform distribution between $0$ and $1$, as in the one--dimensional case. 
At each timestep the species $i$ with lowest fitness and all its $k_i$ neighbours undergo a mutation, and $k_i +1$ new fitness values (drawn from the same uniform distribution) are assigned them. On regular lattices  \cite{BS,BSd}, random graphs  \cite{flyvbjerg}, small--world  \cite{BSsw} and scale--free  \cite{BSsf,BSkim,BSkahng} networks it has been shown that, as for the one--dimensional model, at the stationary state the fitness values are uniformly distributed above a critical threshold $\tau$. The only dependence on the particular topology is the value of $\tau$ \cite{BS,flyvbjerg,BSd, BSsw,BSsf,BSkim,BSkahng}. In particular, $\tau$ vanishes for scale--free degree distributions with diverging second moment \cite{BSsf,BSkim,BSkahng}.\\

While these more complicated networks are closer to realistic food webs   \cite{nature}, as long as the graph is static the model leads to the ecological paradox that, after a mutation, the evolved species inherits the same connections of the previous species. By contrast, macroevolution is believed to be at the same time the cause and the effect of food web dynamics  \cite{webworld}. In particular, after a mutation, a species is expected to develop a new set of interactions with the other species.

\subsection{Definition}
\begin{figure}
\sidecaption
\includegraphics[scale=.45]{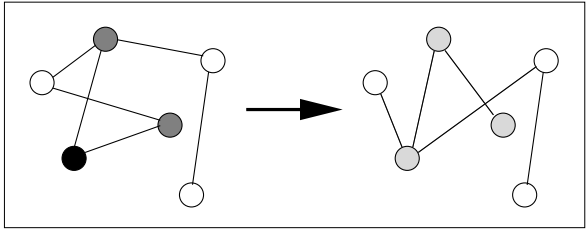}
\caption{Example of graph evolution in the self--organized model. The minimum--fitness vertex (black) and its two neighbours (gray) undergo a mutation: three new fitness values are assigned them (light grey), and new links are drawn  between them and all the other vertices.}
\label{fig:graph}       
\end{figure}

In order to overcome this problem, Garlaschelli et al. assumed that the Bak--Sneppen dynamics is combined with a fitness--driven link updating.
At the initial state the network is generated as in the fitness model, and between all pairs of vertices $i$ and $j$ a link is drawn with probability $f(x_i,x_j)$ (where the $x_i$'s are the initial fitness values).
Then, whenever a species $i$ is assigned a new fitness $x'_i$, all the set of connections between $i$ and the other vertices $j\ne i$ are drawn anew with updated probability $f(x'_i,x_j)$. 
This automatically implies that major mutations (a large change in $x_i$) are associated with very different connection probabilities, while little changes lead to almost equiprobable interactions.
An example of this evolution rule is depicted in figure \ref{fig:graph}.

Two possible choices for updating the fitness of a mutating vertex where proposed. In the original paper  \cite{natphys}, the usual prescription was adopted: each neighbour $j$ of the minimum--fitness vertex receives a fitness drawn anew from the uniform distribution on the unit interval. This means
\begin{equation}
x_j(t+1)=\eta
\end{equation}
where $\eta$ is uniformly distributed between $0$ and $1$. Therefore, $x_j$ is completely updated, independently of its degree $k_j$. 
In another study  \cite{statphys}, a weaker rule was assumed. In particular, the fitness of each neighbour $j$ is assumed to change only by an amount proportional to $1/k_j$:
\begin{equation}
x_j(t+1)=\frac {1}{k_j}\eta + \frac{k_j-1}{k_j}x_j(t)
\label{eq:cicco}
\end{equation}
where again $\eta$ is a random number uniformly distributed between $0$ and $1$. 
Under this second assumption, $x_j$ is completely modified if the only neighbour of $j$ is the minimum--fitness vertex, in which case $k_j=1$. If $j$ has $k_j-1$ additional neighbours, a share $(k_j-1)/k_j$ of $x_j$ is unchanged, and the remaining fraction $x_j/k_j$ is updated to $\eta/k_j$. This makes hubs affected less than small--degree vertices. Clearly, it also implies that the probability of connection to all other vertices varies by a smaller amount. In what follows we shall present both analytical and numerical results derived under the first choice \cite{natphys}. Numerical simulations of the model under the second rule are reported in \cite{statphys}.

\subsection{Analytical solution}
Remarkably, the model is exactly solvable for any choice of the connection probability $f(x,y)$  \cite{natphys}.
Indeed, one can write down a master equation for the fitness distribution $\rho(x,t)$ at time $t$: 
\begin{equation}
\frac{\partial\rho(x,t)}{\partial t}=r^{in}(x,t)-r^{out}(x,t)
\end{equation}
where $r^{in}(x,t)$ and $r^{out}(x,t)$ are the fractions of vertices with fitness $x$ entering and exiting the system at time $t$ respectively. 
If a stationary distribution (time--independent) distribution $\rho(x)$ exists, it is found by requiring
\begin{equation}\label{eqq}
\frac{\partial\rho(x,t)}{\partial t}=0\quad\Rightarrow\quad r^{in}(x)=r^{out}(x)
\end{equation}
where at the stationary state the quantities no longer depend on time. If one manages to write down $r^{in}(x)$ and $r^{out}(x)$ in terms of $f(x,y)$ and $\rho(x)$, then the above condition will give the stationary form of $\rho(x)$ for any choice of $f(x,y)$.\\

To this end, it is useful to introduce the distribution $q(m)$ of the minimum fitness $m\equiv x_{min}$. 
For $x$ small enough, $\rho(x)$ must be very close to $q(x)/N$ (the distribution of all fitness values must be approximated by the correctly renormalized distribution of the minimum).
The range where $\rho(x)\approx q(x)/N$ holds can be defined more formally by introducing the fitness value $\tau$ such that
\begin{equation}
\lim_{N\to\infty}\frac{N\rho(x)}{q(x)}\left\{\begin{array}{ll}
=1&\qquad x\le\tau\\
>1&\qquad x>\tau
\end{array}\right.
\label{eq_limit}
\end{equation}
This means that in the large size limit the fitness distribution for $x<\tau$ is determined by the distribution of the minimum. After an expression for $\rho(x)$ is derived, the value of $\tau$ can be determined by the normalization condition 
\begin{equation}
\int_0^1\rho(x)dx=1
\label{eq_normalization}
\end{equation}
as we show below.
Note that we are not assuming from the beginning that $\tau>0$ as is observed for the Bak--Sneppen model on other networks. It may well be that for a particular choice of $f(x,y)$ eq.(\ref{eq_normalization}) yields $\tau=0$, signalling the absence of a nonzero threshold. 
Also, note that $\lim_{N\to\infty}q(x)=0$ for $x>\tau$, since eq.(\ref{eq_limit}) implies that the minimum is surely below $\tau$. Thus the normalization condition for $q(x)$ reads $\int_0^\tau q(x)dx=1$ as $N\to\infty$.\\

The knowledge of $q(m)$ allows one to rewrite $r^{in}(x)$ and $r^{out}(x)$ as $r^{in}(x)=\int q(m)r^{in}(x|m)dm$ and $r^{out}(x)=\int q(m)r^{out}(x|m)dm$, where $r^{in}(x|m)$, $r^{out}(x|m)$ are conditional probabilities corresponding to the  fractions of vertices with fitness $x$ which are added and removed when the value of the minimum fitness is $m$. 
Let us consider $r^{in}(x)$ first. 
If the minimum fitness is $m$, then $1+k(m)$ new fitness values are updated, where $k(m)$ is the expected degree of the minimum--fitness vertex. Since each of these $1+k(m)$ values is uniformly drawn between $0$ and $1$, one has
\begin{equation}
r^{in}(x|m)=\frac{1+k(m)}{N}
\label{eq_rinm}
\end{equation}
independently of $x$. This directly implies
\begin{equation}
r^{in}(x)=\int_0^\tau q(m)r^{in}(x|m)dm=\frac{1+\langle k_{min}\rangle}{N}
\label{eq_rin}
\end{equation}
where $\langle k_{min}\rangle\equiv\int_0^\tau q(m)k(m)dm$ is the average degree of the vertex with minimum fitness, a quantity that can be derived independently of $k(m)$ as we show below. 
Now consider $r^{out}(x)$, for which the independence on $x$ does not hold. For $x<\tau$, $r^{out}(x|m)=1/N$ if $x=m$ since the minimum is surely replaced. For $x>\tau$, 
the fraction of vertices with fitness $x$ that are removed equals $\rho(x)$ times the probability that a vertex with fitness $x$ is connected to the vertex with minimum fitness $m$.
This probability depends on the fitness values $x'$ and $m'$ that the vertices currently having fitness $x$ and $m$ had at the most recent update of the link connecting them, and simply equals $f(x',m')$  \cite{natphys}.
This means
\begin{equation}
r^{out}(x|m)=\Theta(\tau-x)\frac{\delta(x-m)}{N}+\Theta(x-\tau)\rho(x)f(x,m)
\label{eq_routm}
\end{equation}
where $\Theta(x)=1$ if $x>0$ and $\Theta(x)=0$ otherwise, and $\delta(x)$ is the Dirac delta function. 
An integration over $q(m)dm$ yields
\begin{eqnarray}
r^{out}(x)&=&\int_0^\tau q(m)r^{in}(x|m)dm\nonumber\\
&=&\left\{\begin{array}{ll}
q(x)/N&\quad x<\tau\\
\rho(x)\int_0^\tau q(m)f(x,m)dm&\quad x>\tau
\end{array}\right.
\label{eq_rout}
\end{eqnarray}\\

Finally, one can impose eq.(\ref{eqq}) at the stationary state. If $x<\tau$, this yields $q(x)=1+\langle k_{min}\rangle$ independently of $x$. 
Combining this result with $q(x)=0$ for $x>\tau$ as $N\to\infty$, one finds that the distribution of the minimum fitness $m$ is uniform between $0$ and $\tau$:
\begin{equation}
q(m)=(1+\langle k_{min}\rangle)\Theta(\tau-m) 
\end{equation}
Requiring that $q(m)$ is normalized yields
\begin{equation}
\langle k_{min}\rangle=\frac{1-\tau}{\tau}
\label{eq_kmin}
\end{equation}
Therefore eq.(\ref{eq_rin}) can be written as
\begin{equation}
r^{in}(x)=\frac{1}{\tau N}\qquad\forall x
\end{equation}
If $x>\tau$, eq.(\ref{eqq}) implies
\begin{eqnarray}
\rho(x)&=&\frac{r^{out}(x)}{\int_0^\tau q(m)f(x,m)dm}\nonumber\\
&=&\frac{r^{in}(x)}{\int_0^\tau q(m)f(x,m)dm}\nonumber\\
&=&\frac{1}{\tau N\int_0^\tau q(m)f(x,m)dm}\nonumber\\
&=&\frac{1}{N\int_0^\tau f(x,m)dm}
\end{eqnarray}
which must be equal to $\rho(x)=q(x)/N=(\tau N)^{-1}$ for $x<\tau$. Using this relation, the exact solution for $\rho(x)$ at the stationary state is found  \cite{natphys}:
\begin{equation}\label{eq:solution}
\rho(x)=\left\{\begin{array}{ll}
(\tau N)^{-1} &\quad x<\tau\\
\displaystyle{\frac{1}{N\int_0^\tau f(x,m)dm}}
&\quad x>\tau
\end{array}\right.
\end{equation}
where $\tau$ is determined using eq.(\ref{eq_normalization}), that reads
\begin{equation}
\int_\tau^1\displaystyle{\frac{dx}{\int_0^\tau f(x,m)dm}}=N-1
\label{eq_norm}
\end{equation}
The above analytical solution holds for any form of $f(x,y)$.
As a strikingly novel result, one finds that $\rho(x)$ is in general no longer uniform for $x>\tau$. This unexpected result, which contrasts with the outcomes of the Bak--Sneppen model on any static network, is solely due to the feedback between topology and dynamics. 
At the stationary state the fitness values and the network topology continue to evolve, but the knowledge of $\rho(x)$ allows to compute the expected topological properties as shown in section \ref{sec:fm} for the static fitness model. 

\subsection{Particular cases}
In what follows we consider specific choices of the connection probability $f(x,y)$. In particular, we consider two forms already presented in section \ref{sec:fm}. Once a choice for $f(x,y)$ is made, one can also confirm the theoretical results with numerical simulations. As we show below, the agreement is excellent.

\subsubsection{The random neighbour model}
As we have noted, the trivial choice for the fitness model is $f(x,y)=p$, which is equivalent to the random graph model. 
When the Bak--Sneppen dynamics takes place on the network, this choice removes the feedback with the topology, since the evolution of the fitness does not influences the connection probability. Indeed, this choice is asymptotically equivalent to the so--called \emph{random neighbour} variant  \cite{flyvbjerg} of the Bak--Sneppen model. In this variant each vertex has exactly $d$ neighbours, which are uniformly chosen anew at each timestep. Here, we know that for a random graph the degree is well peaked about the average value $p(N-1)$ (see section \ref{sec:rg}), thus we expect to recover the same results found for $d=p(N-1)$ in the random neighbour model.
Indeed, eq.(\ref{eq:solution}) leads to 
\begin{equation}
\rho(x)=\left\{\begin{array}{ll}
(\tau N)^{-1} &\quad x<\tau\\
(p\tau N)^{-1} &\quad x>\tau
\end{array}\right.
\label{eq_rhop}
\end{equation}
and eq.(\ref{eq_norm}) yields 
\begin{equation}
\tau=\frac{1}{1+pN}\to
\left\{\begin{array}{lll}
1&\quad pN\to 0\\
(1+d)^{-1}&\quad pN=d\\
0&\quad pN\to \infty
\end{array}\right.
\label{eq_tau0}
\end{equation}
The reason for the onset of these three dynamical regimes must be searched for in the topological phases of the underlying network. For $p$ large, there is one large connected component that spans almost all vertices. As $p$ decreases, this \emph{giant cluster} becomes smaller, and several separate clusters form. Below the critical \emph{percolation threshold} $p_c\approx 1/N$  \cite{siam,AB01}, the graph is split into many small clusters. Exactly at the percolation threshold $p_c$, the sizes of clusters are power--law distributed according to $P(s) \propto s^{-\alpha}$ with $\alpha=2.5$  \cite{siam}.
Here we find that the dense regime $pN\to \infty$ is qualitatively similar to a complete graph, where many fitness values are continuously updated and therefore $\tau\to 0$ as in the initial state (thus $\rho(x)$ is not step--like). 
In the sparse case where $pN=d$ with finite $d>1$ as $N\to\infty$, then each vertex has a finite number of neighbours exactly as in the random neighbour model, and one correctly recovers the finite value $\tau=(1+d)^{-1}$ found in ref. \cite{flyvbjerg}. 
The subcritical case when $p$ falls faster than $1/N$ yields a fragmented graph  below the percolation threshold. This is qualitatively similar to a set of $N$ isolated vertices, for which $\tau\to 1$. It is instructive to notice from eq.(\ref{eq:solution}) that the choice $f(x,y)=p$ is the only one for which $\rho(x)$ is still uniform. This confirms that, as soon as the feedback is removed, the novel effects disappear.

\subsubsection{The self--organized configuration model}
\begin{figure}[t]
\sidecaption[t]
\includegraphics[scale=.42]{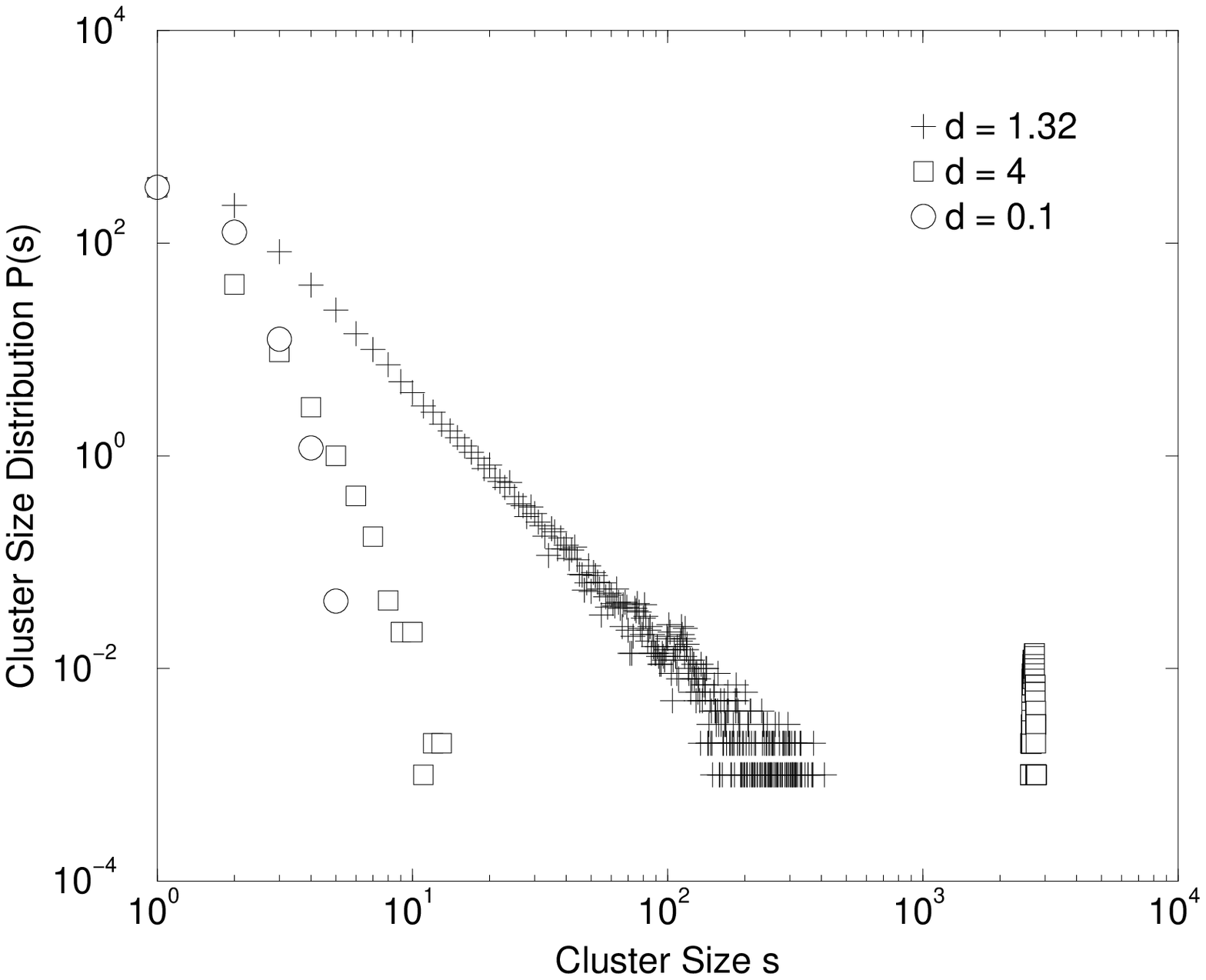}
\caption{Cluster size distribution. Far from the critical threshold 
($d=0.1$ and $d=4$), $P(s)$ is well peaked. At $d_c=1.32$, $P(s)\propto s^{-\alpha}$ with $\alpha=2.45\pm 0.05$. Here $N=3200$. (After ref. \cite{natphys}).}
\label{cluster_size_distribution}
\end{figure}

\begin{figure}[t]
\sidecaption[t]
\includegraphics[scale=.40]{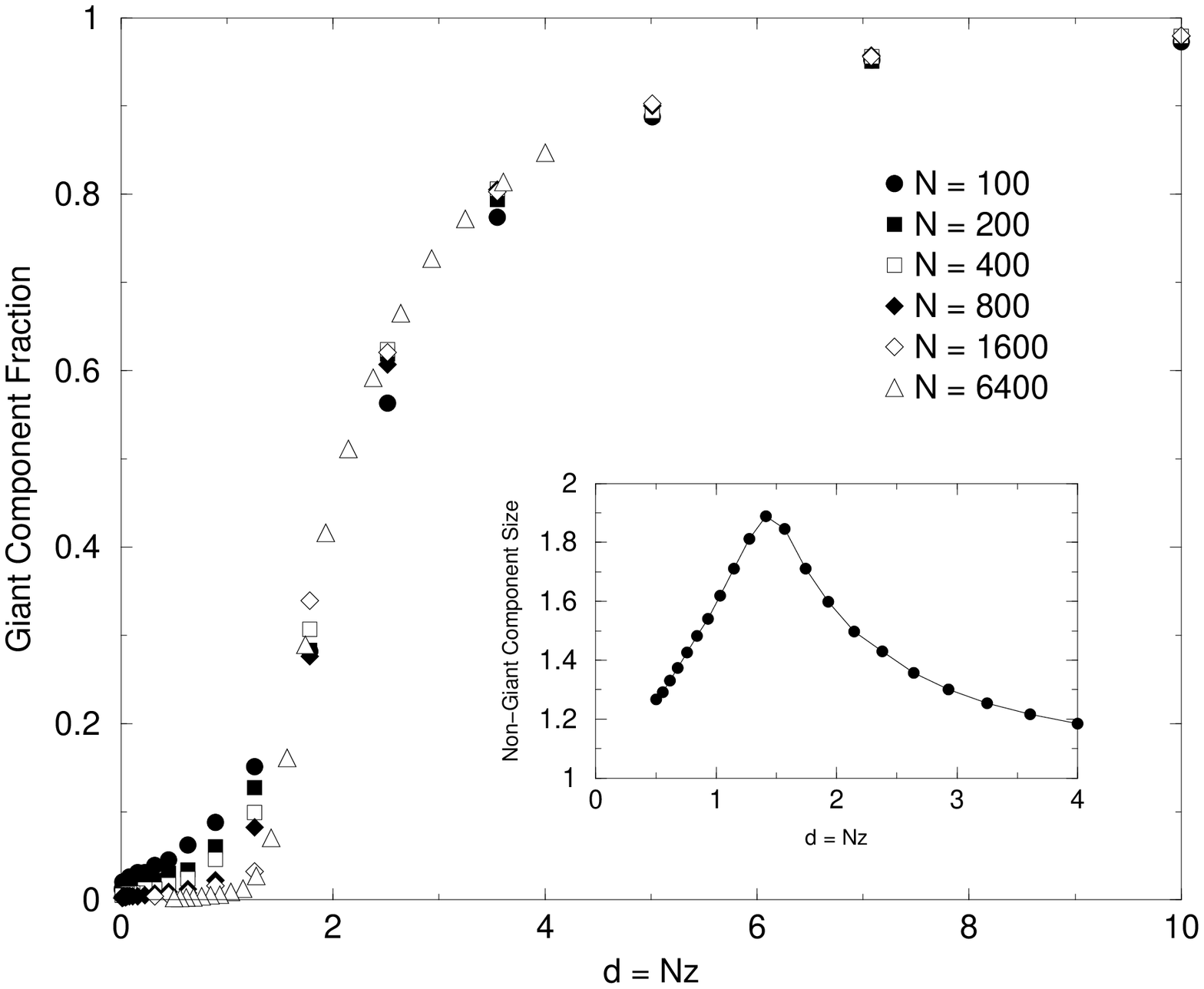}
\caption{Main panel: the fraction of nodes in the giant component for different
network sizes as a function of $d$. Inset: the non-giant component average size as a function of $d$ for $N=6400$. (After ref. \cite{natphys}).}
\label{giant_component}
\end{figure}

Following the considerations in section \ref{sec:fm}, the simplest nontrivial choice for $f(x,y)$ is given by eq.(\ref{fermi}).
For a fixed $\rho(x)$, this choice generates a fitness--dependent version of the \emph{configuration model}  \cite{siam,maslov}, where all graphs with the same degree sequence are equiprobable. All higher--order properties besides the structural correlations induced by the degree sequence are completely random  \cite{newman_origin,likelihood}.
In this self--organized case, the degree sequence is not specified \emph{a priori} and is determined by the fitness distribution at the stationary state. 
Inserting eq.(\ref{fermi}) into eq.(\ref{eq:solution}) one finds a solution that for $N\to\infty$ is equivalent to  \cite{natphys}
\begin{equation}
\rho(x)=\left\{\begin{array}{ll}
(\tau N)^{-1} &\quad x<\tau\\
(\tau N)^{-1}+2/(zN\tau^2x) &\quad x>\tau
\end{array}\right.
\label{eq_rhofermi}
\end{equation}
where $\tau$, again obtained using eq.(\ref{eq_norm}), is
\begin{equation}
\tau=\sqrt{\frac{\phi(zN)}{zN}}\to
\left\{\begin{array}{lll}
1&\quad zN\to 0\\
\sqrt{\phi(d)/d}&\quad zN=d\\
0&\quad zN\to \infty
\end{array}\right.
\label{eq_taufermi}
\end{equation}
Here $\phi(x)$ denotes the ProductLog function, defined as the solution of $\phi e^\phi=x$. Again, the above dynamical regimes are related to three 
(subcritical, sparse and dense) underlying topological phases. This can be ascertained by monitoring the cluster size distribution $P(s)$. It is found that $P(s)$ develops a power--law shape $P(s) \propto s^{-\alpha}$ (with $\alpha=2.45 \pm 0.05$) when $d\equiv zN$ is set to the critical value $d_c =1.32 \pm 0.05$  \cite{natphys} (see fig. \ref{cluster_size_distribution}), which therefore represents the percolation threshold. 
This behaviour can also be explored by measuring the fraction of vertices spanned by the giant cluster as a function of $d$ (see fig. \ref{giant_component}). This quantity is negligible for
$d<d_c$, while for $d>d_c$ it takes increasing finite values. Also, one can plot the average size fraction of non--giant components. As shown in the
inset of fig. \ref{giant_component}, this quantity diverges at the critical point where $P(s)$ is a power law. 
\\

\begin{figure}[t]
\includegraphics[scale=.9]{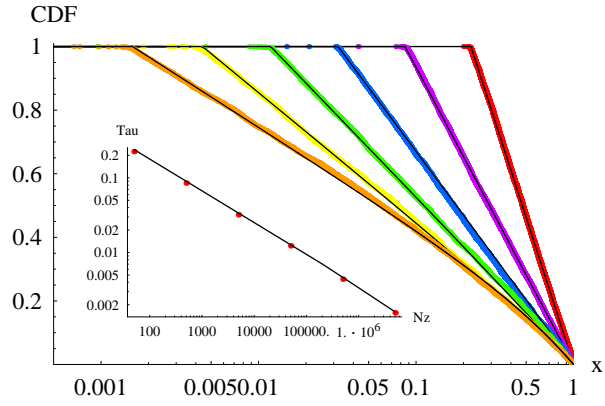}
\caption{
Main panel: cumulative density function $\rho_>(x)$ in log--linear axes. From right to left, $z = 0.01$, $z = 0.1$, $z = 1$, $z = 10$, $z = 100$, $z=1000$ ($N=5000$).  Inset: log--log plot of $\tau(zN)$. Solid lines: theoretical curves, points: simulation results. (After ref. \cite{natphys}).}
\label{fig_px}
\end{figure}

The analytical results in eq.(\ref{eq_rhofermi}) mean that $\rho(x)$ is the superposition of a uniform distribution and a power--law with exponent $-1$.
The decay of $\rho(x)$ for $x>\tau$ is entirely due to the coupling between extremal dynamics and topological restructuring. It originates from the fact that at any time the fittest species is also the most likely to be selected for mutation, since it has the largest probability to be connected to the least fit species. This is opposite to what happens on fixed networks.
The theoretical predictions in eqs.(\ref{eq_rhofermi}) and (\ref{eq_taufermi}) can be confirmed by large numerical simulations. 
This is shown in fig.\ref{fig_px}, where the cumulative fitness distribution $\rho_>(x)$ defined in eq.(\ref{eq:rho>}) and the behaviour of $\tau(zN)$ are plotted. 
Indeed, the simulations are in very good accordance with the analytical solution. 
Note that, as we have discussed in section \ref{sec:fm}, in the sparse regime $z\ll 1$ one has $f(x,y)\approx zxy$. Here, this implies a purely power--law behaviour $\rho(x)\propto x^{-1}$ for $x>\tau$. Therefore $\rho_>(x)$ is a logarithmic curve that looks like a straight line in log--linear axes. 
In the dense regime obtained for large $z$, the uniform part gives instead a significant deviation from the power--law trend. This shows one effect of structural correlations.
\\

\begin{figure}[t]
\includegraphics[scale=1.1]{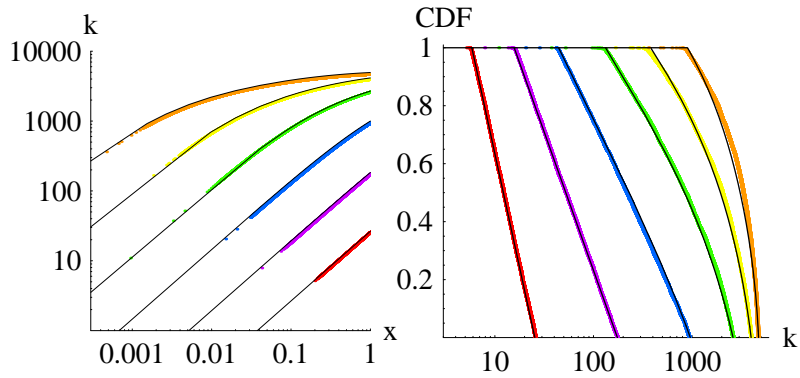}
\caption{Left: $k(x)$ ($N=5000$; from right to left, $z=0.01$, $z=0.1$, $z=1$, $z=10$, $z=100$, $z=1000$). Right: $P_>(k)$ (same parameter values, inverse order from left to right). Solid lines: theoretical curves, points: simulation results. (After ref. \cite{natphys}).}
\label{fig4}
\end{figure}

Other effects are evident when considering the degree distribution $P(k)$. 
Using eq.(\ref{eq:kx}) one can obtain the analytic expression of the expected degree $k(x)$ of a vertex with fitness $x$:
\begin{equation}
k(x)=\frac{2}{z\tau^2}\ln\frac{1+zx}{1+z\tau x}+\frac{zx-\ln(1+zx)}{z\tau x}
\label{eq_k}
\end{equation}
Computing the inverse function $x(k)$ and plugging it into eq.(\ref{eq:Prho}) allows to obtain the cumulative degree distribution $P_>(k)$. 
Both quantities are shown in fig.\ref{fig4}, and again the agreement between theory and simulations is excellent. For small $z$, $k(x)$ is linear, while for large $z$ a saturation to the maximum value $k_{max} = k(1)$ takes place.
As discussed in section \ref{sec:fm}, this implies that in the sparse regime $P(k)$ has the same shape as $\rho(x)$. 
Another difference from static networks is that here $\tau$ remains finite even if $P(k)\propto k^{-\gamma}$ with $\gamma<3$  \cite{BSsf,BSkim,BSkahng}. 
For large $z$ the presence of structural correlations introduces a sharp cut--off for $P(k)$. 

\section{Conclusions}
We have presented a brief, and by no means complete, summary of the ideas that inspired much of the research on scale--invariance and self--similarity, from the early discovery of fractal behaviour to the more recent study of scale--free networks. We have highlighted the importance of understanding the emergence of the ubiquitously observed patterns in terms of dynamical models. In particular, the framework of Self--Organized Criticality succeeds in explaining the onset of fractal behaviour without external fine--tuning. According to the SOC paradigm, open dissipative systems appear to evolve spontaneously to a state where the response to an infinitesimal perturbation is characterized by avalanches of all sizes. We have emphasized the importance of introducing similar mechanisms in the study of networks. In particular, we have argued that in many cases of interest it is not justified to decouple the formation of a network from the dynamics taking place on it. In both cases, one is forced to introduce \emph{ad hoc} specifications for the process assumed to be slower. Indeed, by presenting an extensive study of a self--organized network model, we have shown that if the feedback between topology and dynamics is restored, novel and unpredictable results are found. 
This indicates that adaptive networks provide a more complete explanation for the spontaneous emergence of complex topological properties in real networks.


\input{referenc}

\end{document}

%% file: referenc.tex
%
%
%